\documentclass[prl,aps,twocolumn,groupedaddress,showpacs,floatfix]{revtex4}
\usepackage{amsmath,amssymb,multirow,epsfig,bm}

\newcommand{\beq}{\begin{equation}}
\newcommand{\eeq}{\end{equation}}
\newcommand{\bea}{\begin{eqnarray}}
\newcommand{\eea}{\end{eqnarray}}

\begin{document}
\title {Is graphene in vacuum an insulator?}

\author{Joaqu\'{\i}n E. Drut$^1$ and Timo A. L\"ahde$^2$ }
\affiliation{$^1$Department of Physics, The Ohio State University, Columbus, OH 43210--1117, USA}
\affiliation{$^2$Department of Physics, University of Washington, Seattle, WA 98195--1560, USA}

\begin{abstract}

We present evidence, from Lattice Monte Carlo simulations of the phase diagram of graphene as a function of the Coulomb 
coupling between quasiparticles, that graphene in vacuum is likely to be an insulator. We find a semimetal-insulator 
transition at $\alpha_g^\text{crit} = 1.11 \pm 0.06$, where $\alpha_g^{} \simeq 2.16$ in vacuum, and $\alpha_g^{} \simeq 
0.79$ on a SiO$_2^{}$ substrate. Our analysis uses the logarithmic derivative of the order parameter, supplemented by 
an equation of state. The insulating phase disappears above a critical number of four-component fermion flavors 
$4 < N_f^{\text{crit}} < 6$. Our data are consistent with a second-order transition.


\end{abstract}

\date{\today}

\pacs{73.63.Bd, 71.30.+h, 05.10.Ln}
\maketitle


Graphene, a carbon allotrope with a two-dimensional honeycomb structure, has become an important player at the forefront of 
condensed matter physics, drawing the attention of theorists and experimentalists alike due to its challenging nature as a 
many-body problem, its unusual electronic properties and possible technological applications 
(see~Refs.~\cite{CastroNetoetal,GeimNovoselov} and references therein). Graphene also belongs to a large class of planar 
condensed-matter systems, which includes other graphite-related materials as well as high-$T_c^{}$ superconductors.

A distinctive feature of graphene is that its band structure contains two degenerate `Dirac points', in 
the vicinity of which the dispersion is linear, as in relativistic theories~\cite{Semenoff}. 
The low-energy excitations in graphene are thus Dirac quasiparticles of Fermi velocity $v \simeq c/300$, 
where $c$ is the speed of light in vacuum. These are described by the Euclidean action
\begin{eqnarray}
S_E^{} &=& -\sum_{a=1}^{N_f^{}} \int d^2x\,dt \: \bar\psi_a^{} \:D[A_0^{}]\: \psi_a^{} 
\nonumber \\
&& +\,\frac{1}{2g^2} \int d^3x\,dt \: (\partial_i^{} A_0^{})^2,
\label{SE}
\end{eqnarray}
where $g^2 = e^2/\epsilon_0^{}$ for graphene in vacuum, $\psi_a^{}$ is a four-component Dirac field in 2+1~dimensions, 
$A_0^{}$ is a Coulomb field in 3+1 dimensions, $N_f^{}=2$ for real graphene, and
\begin{eqnarray}
D[A_0^{}] &=& \gamma_0^{} (\partial_0^{} + iA_0^{}) + v\gamma_i^{} \partial_i^{}, \quad i=1,2
\end{eqnarray}
where the Dirac matrices $\gamma_\mu^{}$ satisfy the Euclidean Clifford algebra $\{ \gamma_\mu^{}, \gamma_\nu^{}\} = 
2\delta_{\mu\nu}^{}$. The strength of the Coulomb interaction is controlled (as can be shown by rescaling $t$ and 
$A_0^{}$) by $\alpha_g^{} = e^2/(4 \pi v \epsilon_0^{})$, which is the graphene analogue of the fine-structure constant 
$\alpha \simeq 1/137$ of quantum electrodynamics~(QED).



Despite the similarities with QED, the smallness of $v/c$ in graphene has non-trivial consequences: Coulomb interactions 
between the quasiparticles are essentially instantaneous, thus breaking relativistic invariance which is reflected in 
Eq.~(\ref{SE}). The analogue of the fine-structure constant $\alpha_g^{} \simeq 300\,\alpha$ in graphene, such that the 
low-energy properties resemble~QED in a very strongly coupled regime. This provides an exciting opportunity for the study of 
strongly coupled theories, within a condensed-matter analogue that can be experimentally realized with modest equipment.

Notably, Eq.~(\ref{SE}) satisfies a chiral U($2N_f^{}$) symmetry which can break spontaneously at large enough Coulomb 
coupling, generating a gap in the quasiparticle spectrum. Whether such an effect occurs in 
real graphene is an open issue from the experimental point of view (see however Ref.~\cite{exp}, where a substrate-induced 
gap is reported). On the theoretical side, dynamical gap generation is described by a quantum phase transition due to the 
formation of particle-hole bound states. However, in such a strongly coupled regime, even a qualitative analysis should be 
non-perturbative. This is especially true for graphene in vacuum, where $\alpha_g^{}$ attains its maximum value, while it is 
partially screened in the presence of a substrate. While the semimetallic properties of graphene on a substrate are well 
established, the issue of whether a transition to an insulating phase occurs in the absence of a substrate remains 
unsettled.

This problem has been studied using perturbative as well as non-perturbative methods~\cite{Gonzalez, 
Wagner,Leal:2003sg,Miransky}. The latter, which are typically based on a gap equation, yield an infinite-order 
transition to an insulating phase above a critical coupling. On the other hand, the results of large-$N_f^{}$ 
analyses~\cite{Gonzalez,Herbut,Son} find that Coulomb interactions flow towards a non-interacting fixed point 
under renormalization-group~(RG) transformations, being therefore unable to induce a transition. However, a number of 
uncontrolled approximations are involved, such as the reliance on large-$N_f^{}$ results that may break down for small 
$N_f^{}$, various approximate treatments of the gap equation kernel, as well as the linearization of the resulting integral 
equation (see Appendix~B of Ref.~\cite{Miransky}). 

We set out to characterize the phase diagram of graphene in a controlled fashion, which 
entails a lattice Monte Carlo approach and analysis of the chiral condensate, which is the order parameter for a transition 
into an insulating charge-density-wave phase. Such an approach is non-perturbative, takes full account of quantum 
fluctuations, and has been extensively used~\cite{Kogutetal,Gockeleretal,AliKhan:1995wq,Hands:2002dv} in the study of 
2+1~(QED$_3^{}$) and 3+1~(QED$_4^{}$) dimensional QED, but not for studies of graphene (see however 
Ref.~\cite{HandsStrouthos}, where a model for the strong-coupling limit is investigated).

To this end, we discretize the pure gauge part of 
Eq.~(\ref{SE}) according to (for a recent overview, see Ref.~\cite{Rothe})
\begin{eqnarray}
S^g_E[\theta_0^{}] &=& \frac{\beta}{2} \sum_{\bm{n}}
{\left[\sum^3_{i=1} \left(\theta_{0,\bm{n}}^{} - \theta_{0,\bm{n} + \bm{e}_i^{}}^{}\right)^2_{} \right]},
\label{Sg}
\end{eqnarray}
where the (dimensionless) lattice coupling $\beta \equiv v/g^2$, $\theta_0^{}$ is the lattice gauge potential, 
$\bm n \equiv (n_0^{},\ldots,n_3^{})$ denotes a site on the space-time lattice, 
and ${\bm e}_\mu^{}$ is a unit vector in the direction $\mu$. 
For studies of chiral phase transitions, staggered fermions~\cite{Kogut-Susskind} are a preferred choice, since 
chiral symmetry is then partially preserved. As $N$ staggered flavors correspond to $N_f^{} = 2N$ continuum Dirac 
flavors~\cite{BurdenBurkitt}, it suffices (for real graphene) to set $N=1$, which gives
\begin{eqnarray}
S^f_E[\bar{\chi},\chi,U_0^{}] &=& 
-\sum_{\bm{m},\bm{n}}
\bar\chi_{\bm{m}}^{} \: D_{\bm{m},\bm{n}}^{}[U_0^{}] \: \chi_{\bm{n}}^{},
\label{Sf}
\end{eqnarray}
%
%
%
where the $\chi_{\bm{n}}^{}$ are staggered fermion spinors, and $(\bm{m},\bm{n})$ are restricted to 
a 2+1~dimensional sublattice. The invariance of Eq.~(\ref{SE}) under spatially uniform, time-dependent gauge 
transformations is retained by coupling the fermions to the gauge field via $U_0^{} = \exp(i\theta_0^{})$. 
The staggered form of $D$ is
\begin{eqnarray}
D_{\bm{m},\bm{n}}^{}[U_0^{}] &\!\!=\!\!&
\frac{1}{2} \left[\delta_{\bm{m}+\bm{e}_0^{},\bm{n}}^{}\,U_{0,\bm{m}}^{} 
- \delta_{\bm{m}-\bm{e}_0^{},\bm{n}}^{}\,U_{0,\bm{n}}^{\dagger}\right]
\label{Df} \\
&\!\!+\!\!&\frac{1}{2} \sum_i \eta_{i,\bm{m}}^{} \left[\delta_{\bm{m}+\bm{e}_i^{},\bm{n}}^{} 
- \delta_{\bm{m}-\bm{e}_i^{},\bm{n}}^{}\right] 
+ m_0^{}\,\delta_{\bm{m},\bm{n}}^{} \nonumber 
\end{eqnarray}
%
%
%
where $\eta_{1,\bm{n}}^{} = (-1)^{n_0^{}}$ and $\eta_{2,\bm{n}}^{} = (-1)^{n_0^{} + n_1^{}}$.
The mass term breaks chiral symmetry explicitly, generating a non-zero condensate which is otherwise not possible at 
finite volume. Extrapolation to $m_0^{}=0$ is thus required. Upon integration of the fermionic degrees of freedom, 
the path integral is governed by the effective action
\begin{eqnarray}
S_{\text{eff}}[\theta_0^{}] = -N\ln\det(D_{\bm{m},\bm{n}}^{}[U_0^{}]) + S^g_E[\theta_0^{}],
\label{Seff}
\end{eqnarray}
such that $P[\theta_0^{}] \equiv \exp(-S_{\text{eff}}[\theta_0^{}])$ defines the Monte Carlo probability measure.
It is straightforward to show that the determinant is positive definite. 
We have sampled $P[\theta_0^{}]$ using the Metropolis algorithm, updating $\theta_0^{}$ at random locations
and evaluating the fermion determinant exactly. Our approach has been tested against known results for QED$_3^{}$ and 
QED$_4^{}$. The data of Ref.~\cite{Kogutetal} on the chiral condensate of QED$_3^{}$ have been accurately 
reproduced for multiple values of $N_f^{}$, along with several randomly chosen datapoints from 
Refs.~\cite{Gockeleretal,Hands:2002dv}.


%
\begin{figure}[t]
\epsfig{file=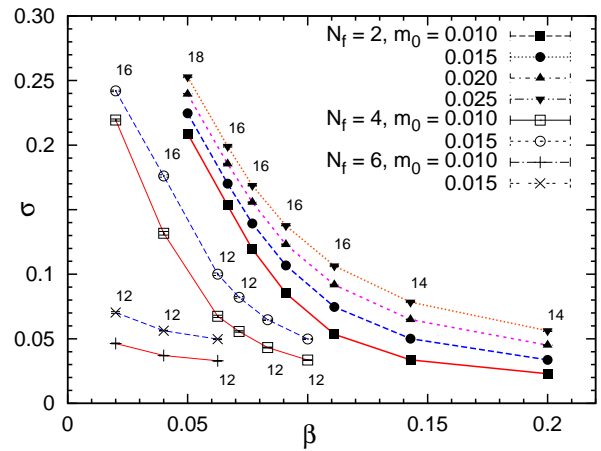, width=\columnwidth}
\caption{(Color online) 
Chiral condensate $\sigma$ for $N_f^{} = 2,4,6$ as a function of $\beta$ and $m_0^{}$, with lines intended to guide the 
eye. The lattices are of extent $L^3_{}\times L_z^{}$, such that the fermions live in a 2+1~dimensional cube of size 
$L$, while the gauge bosons also propagate in the $z$-direction of length $L_z^{}$. For each $\beta$, the value of $L$ is 
given next to the datapoints. All results are for $L_z^{}\! = 8$, as larger values had no discernible 
effects. For each datapoint $\sim 300$ uncorrelated gauge configurations were generated. The statistical 
uncertainties, which are comparable to the size of the symbols, were obtained by the jackknife method~\cite{jackknife}. 
Finite volume effects are largest for small $\beta$. \label{xbarx}}
\end{figure}
The chiral U($2N_f^{}$) symmetry of the continuum theory can only partially be realized on the lattice if the 
doubling problem is to be avoided~\cite{NielsenNinomiya}. In particular, only a global U($N$)$\times$U($N$) symmetry 
remains upon discretization~\cite{BurdenBurkitt}. We focus on the spontaneous breakdown of this symmetry to a U($N$) 
subgroup, characterized by a condensate $\sigma \equiv \langle \bar\chi \chi \rangle \neq 0$ in the limit $m_0^{} \to 0$, 
which marks the appearance of a gap in the quasiparticle spectrum. Our results for $\sigma$ are presented in 
Fig.~\ref{xbarx} for $\beta = 0.05,\ldots,0.5$ and $m_0^{} = 0.010,\ldots,0.025$ (in lattice units). 

\begin{figure}[t]
\epsfig{file=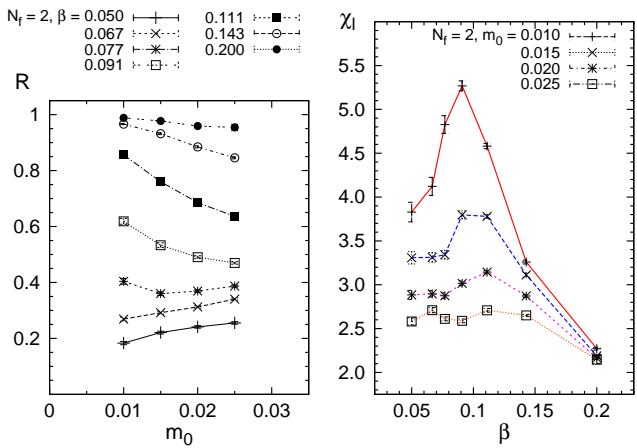, width=\columnwidth}
\caption{\label{Fig_R} (Color online) Left panel: Logarithmic derivative $R$ as a function of $m_0^{}$
for different $\beta$. Right panel: Chiral susceptibility $\chi_l^{}$ as a function of $\beta$ for 
different $m_0^{}$. All data are for $N_f^{} = 2$, with lattice sizes identical to those of Fig.~\ref{xbarx}.
The lines are intended as a guide to the eye.} 
\end{figure}

Our data for $N_f^{} = 2$ in Fig.~\ref{xbarx} are suggestive of a critical coupling $\beta_c \sim 0.06 \ldots 0.09$, below 
which $\sigma$ survives in the limit $m_0^{} \to 0$. More significantly, the susceptibility 
$\chi_l^{} = \partial \sigma/\partial m_0^{}$ shown in the right panel of Fig.~\ref{Fig_R} exhibits a maximum which
tends towards $\beta_c^{}$ as $m_0^{}$ is decreased. The (much more limited) data for $N_f^{} = 
4$~(not shown) have a similar maximum around $\beta \sim 0.03$. As $N_f^{}$ is increased, $\sigma$ obviously becomes 
suppressed, vanishing between $N_f^{} = 4$ and $N_f^{} = 6$. This agrees with large-$N_f^{}$ 
results~\cite{Son} that yield a quantum critical point (and therefore no condensate) in the limit $\beta \to 0$, and is 
consistent with recent Monte Carlo studies of that limit~\cite{HandsStrouthos}. Our results for small 
$N_f^{}$ establish that the strong-coupling critical point disappears below $N_f^{{\text{crit}}}$, with $4 < 
N_f^{\text{crit}} < 6$. 

For a quantitative determination of $\beta_c^{}$, we compute the logarithmic derivative $R$ of $\sigma$ with respect to 
$m_0^{}$,
\begin{eqnarray}
R &\equiv & \left.\frac{\partial \ln\sigma}
{\partial \ln m_0^{}}\right|_\beta^{}
= \left.\frac{m_0^{}}\sigma \left(
\frac{\partial \sigma}{\partial m_0^{}}\right)\right|_\beta^{},
\label{Rrat}
\end{eqnarray}
according to the method of Ref.~\cite{Kocic:1992pf}. In the limit $m_0^{} \to 0$, the behavior of $R$ is as follows: $R \to 
1$ in the chirally symmetric (unbroken) phase, where $\sigma \propto m_0^{}$. At the critical coupling $\beta=\beta_c^{}$ 
one finds that $R \to 1/\delta$, where $\delta$ is a universal critical exponent. $R$ vanishes in the 
spontaneously broken phase, where $\sigma \neq 0$ for $m_0^{} \to 0$. The data in Fig.~\ref{Fig_R} (left panel) indicate 
that chiral symmetry is spontaneously broken for $\beta = 0.067$, but remains unbroken for $\beta = 0.077$, from which we 
conclude that $\beta_c^{} = 0.072 \pm 0.005$. This estimate can be refined by use of larger lattice volumes and smaller 
values of $m_0^{}$.

\begin{figure}[b]
\epsfig{file=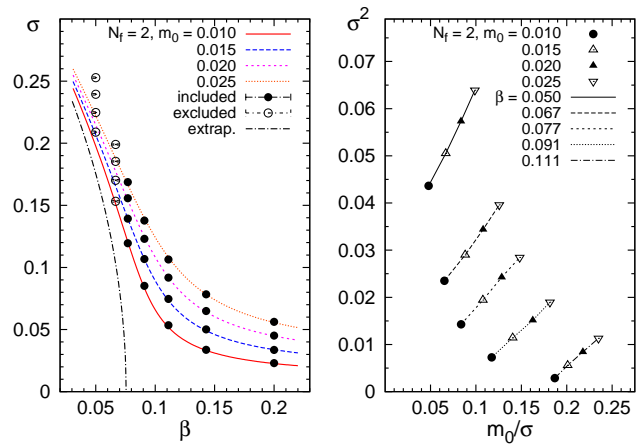, width=\columnwidth}
\caption{(Color online) 
Left panel: $\chi^2_{}$ fit to the data of Fig.~\ref{xbarx} and extrapolation to $m_0^{} = 0$ for 
$N_f^{} = 2$ using Eq.~(\ref{EOS}) with $X_0^{},X_1^{},Y_1^{}$ and $\beta_c^{}$ as free parameters. The points with largest 
finite-volume effects have been excluded from the fit. The optimal parameter values are $\beta_c^{} = 0.0755 \pm 0.0003$, 
$X_0^{} = 0.195 \pm 0.003$, $X_1^{} = -0.089 \pm 0.001$ and $Y_1^{} = -0.091 \pm 0.001$. The uncertainties are 
purely statistical.
Right panel: Fisher plot of $\sigma^2{}$ versus $m_0^{}/\sigma$ for the data of Fig.~\ref{xbarx}
with $N_f^{} = 2$. The lines connect datapoints with identical $\beta$, such that straight lines indicate mean-field 
behavior according to Eq.~(\ref{EOS}). At $\beta_c^{}$, the extrapolation crosses the origin. 
\label{Fitplot}}
\end{figure}

A more precise determination of $\beta_c^{}$ requires an equation of state~(EOS) of the form 
$m_0^{} = f(\sigma,\beta)$ for the extrapolation $m_0^{} \to 0$. We have considered the EOS successfully 
applied~\cite{Gockeleretal,AliKhan:1995wq} to QED$_4^{}$,
\begin{eqnarray}
m_0^{}X(\beta) &=& Y(\beta) f_1^{}(\sigma) + f_3^{}(\sigma),
\label{EOS}
\end{eqnarray}
where $X(\beta)$ and $Y(\beta)$ are expanded around $\beta_c^{}$ such that
$X(\beta) = X_0^{} + X_1^{}(1-\beta/\beta_c^{})$ and $Y(\beta) = Y_1^{}(1-\beta/\beta_c^{})$.
The dependence on $\sigma$ is given by $f_1^{}(\sigma) = \sigma^b_{}$ and $f_3^{}(\sigma) = \sigma^\delta_{}$,
which allows for non-classical critical exponents $\delta$ and $\bar\beta$~\cite{Gockeleretal,AliKhan:1995wq}, 
where $b \equiv \delta - 1/\bar\beta$. A $\chi^2_{}$ fit of Eq.~(\ref{EOS}) to the data in Fig.~\ref{xbarx} is given 
in Fig.~\ref{Fitplot} (left panel), along with a ``Fisher plot'' of $\sigma^2$ versus $m_0^{}/\sigma$ (right panel). 
Deviations from classical mean-field behavior manifest themselves in the Fisher plot as curvature in the lines of constant 
$\beta$. The mostly straight lines in Fig.~\ref{Fitplot} (right panel) indicate that the data should be well described, up 
to finite-size effects, by the classical values $\delta = 3$ and $b=1$, which is confirmed in Fig.~\ref{Fitplot} (left 
panel). 


For all values of $m_0^{}$, the finite-volume effects appear to be dominated by a dynamically generated 
correlation length in the region that we identify as the spontaneously broken phase. To gauge the impact of such effects 
on $\beta_c^{}$, several fits were performed from $L=8$ up to the main fit with $L=16$. Our analysis 
indicates that $\beta_c^{}$ is stable around $\sim 0.075$, even though the datapoints at the smallest $\beta$ 
shift due to finite-volume effects. In contrast, for $N_f^{} = 6$ chiral symmetry remains unbroken, and thus 
finite-volume effects remain small in the limit $\beta \to 0$. The stability of $\beta_c^{}$ can be understood in 
terms of the results for $R$ in Fig.~\ref{Fig_R}, since any fit to the condensate should be consistent with the 
susceptibility as well, and $R$ combines both pieces of information. The character of $R$ clearly changes around $\beta = 
0.075$, in direct correlation with the fitted values of $\beta_c^{}$. If one allows for extreme modifications, such as 
non-classical critical exponents, or forces the fit to account for all the datapoints, values of $\beta_c^{}$ as low as 
$0.060$ may be found, although at the price of a much worse fit to the data. By inclusion and exclusion of different sets of 
datapoints, a realistic (though somewhat model-dependent) estimate of the critical coupling is $\beta_c^{} = 0.0755 \pm 
0.0008$. Possible deviations from classical mean-field behavior with $f_1^{} = \sigma$ and $f_3^{} = \sigma^3_{}$ are 
below the resolution of the present study.


The results in Fig.~\ref{Fitplot} suggest a second-order transition with classical exponents, unlike 
Refs.~\cite{Leal:2003sg,Miransky}, where an infinite-order transition was found. In this situation, further investigation is 
clearly called for. While the sensitivity of our analysis increases for smaller $m_0^{}$, larger lattice volumes are also 
required to keep finite-volume effects under control. A similar EOS~analysis has recently been performed by Hands and 
Strouthos~(Ref.~\cite{HandsStrouthos}) for a graphene-like theory with a zero-range interaction. Unfortunately, a meaningful 
comparison is not possible at this time, as their parameter $1/g^2$ cannot be identified with our gauge coupling $\beta$, 
except in the strong-coupling limit $\beta \rightarrow 0$.


A comparison with experiment necessitates a discussion of renormalized quantities. While test charges remain
unscreened as the fluctuations of the fermion action are confined to $2+1$~dimensions, the physical value of 
$\beta_c^{}$ may be affected by renormalization of $v$ due to the breaking of relativistic invariance. Large-$N_f^{}$ 
results suggest~\cite{Gonzalez,Son} that the Coulomb interaction renormalizes $v$ logarithmically toward larger 
values, thereby decreasing $\alpha_g^{}$ slightly from the bare value, and strengthening our conclusions for graphene in 
vacuum. Available experimental evidence~\cite{exp,Leal:2003sg} indicates that velocity renormalization 
is at most a $\sim 20$\% effect, and of phononic rather than Coulombic origin.

Summarizing, we have found that graphene 
should become insulating at a critical coupling $\alpha_g^{{\text{crit}}} \equiv 1/(4\pi \beta_c^{}) = 1.11 \pm 0.06$, 
where $\beta_c = 0.072 \pm 0.005$. This should be compared with $\alpha_g^{} \simeq 2.16$ in vacuum, and 
$\alpha_g^{} \simeq 0.79$ on an SiO$_2^{}$ substrate (using the experimental value $v\simeq 10^6$~m/s). These findings 
are in line with the observed semimetallic properties~\cite{GeimNovoselov} of graphene on a SiO$_2^{}$ substrate, and 
predict that the Coulomb interaction in suspended graphene should induce a gap in the quasiparticle spectrum. Within the 
accuracy of the present study, the transition appears to be of second order.

Ultimately, the observation of the insulating phase is dependent on the size of the induced band gap. However, the 
prediction of a dimensionful observable requires the matching of a lattice quantity (other than the gap itself) to the 
corresponding experimental value. This applies to finite temperature studies as well, where it is necessary to fix the 
absolute temperature scale. An intriguing possibility is that the observed nanoscale ripples in suspended 
graphene~\cite{Susp} may provide the necessary information, as such corrugations can by described~\cite{Guinea} by means of 
external gauge fields with known dimensionful properties. Exploratory work in this direction, along with a more accurate 
study of the transition properties, is in progress~\cite{DrutLahde2}.


\begin{acknowledgments} 

We acknowledge support under U.S. DOE Grants No.~DE-FG-02-97ER41014, No.~DE-FG02-00ER41132, and No.~DE-AC02-05CH11231, 
UNEDF SciDAC Collaboration Grant No.~DE-FC02-07ER41457 and NSF Grant No.~PHY--0653312. This work was supported
in part by an allocation of computing time from the Ohio Supercomputer Center. We thank A.~Bulgac 
and M.~J.~Savage for computer time, and W.~Detmold, M.~M.~Forbes, R.~J.~Furnstahl, D.~Gazit and D.~T.~Son 
for instructive discussions.

\end{acknowledgments}


\end{document}